\documentclass[epj]{svjour}

\usepackage{amsmath,amssymb,bm}
\usepackage{epsfig}
\usepackage{graphicx}
\usepackage{amsfonts}
\usepackage{epstopdf}
\usepackage[usenames,dvipsnames]{color} 
\usepackage{float}

\newcommand{\eq}{\begin{equation}}
\newcommand{\eeq}{\end{equation}}
\newcommand{\bd}[1]{ \mbox{\boldmath $#1$}  }

\def\ii{\'i}
\newcommand{\beqa}{\begin{eqnarray}}
\newcommand{\eeqa}{\end{eqnarray}}

\begin{document}

\title{
$^{12}$C within the Semimicroscopic Algebraic Cluster Model
}

\author{
P. O. Hess\inst{1,2}
	}

\institute{Instituto de Ciencias Nucleares, Universidad Nacional Aut\'onoma de M\'exico, 
Ciudad Universitaria, Circuito Exterior S/N, A.P. 70-543, 04510
M\'exico D.F. M\'exico
\email{hess@nucleares.unam.mx}
\and Frankfurt Institute for Advanced Studies, Johann Wolfgang Goethe
 Universit\"at, Ruth-Moufang-Str.1, 60438 Frankfurt am Main, Germany  
}

\date{Received: date / Revised version: date}
%
\abstract{
The Semimicroscopic Algebraic Cluster Model (SACM) is applied to $^{12}$C
as a system of three $\alpha$-clusters. The microscopic model space, which observes
the {\it Pauli-Exclusion-Principle} (PEP), is constructed. It is shown that the $^{12}$C nucleus
can effectively be treated as a two-cluster system $^{8}$Be+$\alpha$. The experimental spectrum is well
reproduced.
The geometrical mapping is discussed and it is shown that the ground state
must correspond to a triangular structure, which is in agreement with other microscopic calculations.
The non-zero $B(E2; 0_2^+ \rightarrow 2_1^+)$ transition requires
a mixing of $SU(3)$ {\it irreducible representations} (irreps) whose 
consequences are discussed. The Hoyle state turns out to contain large
shell excitations.
The results are compared to another phenomenological model, which
assumes a triangular structure and, using simple symmetry arguments, can reproduce
the states observed at low energy. This model does not observe the PEP and one
objective of our contribution is to verify the extend of importance of the PEP.
} 

\maketitle
\section{Introduction}
\label{intro}

In recent years,
the $^{12}$C nucleus enjoyed an increased interest, especially 
concerning the structure of the Hoyle state \cite{hoyle}. 
Experiments revealed new states
\cite{bijker} and microscopic calculations show a triangular structure with peaks in  the density
corresponding approximately to the center of $\alpha$ particles. Particularly, the Hoyle
state \cite{hoyle} is important in the understanding of the fusion in stars of three $\alpha$-particles
to the $^{12}$C nucleus. In \cite{draayer-hoyle} a no-core shell model calculation was 
performed in order to describe this state, requiring many shell excitations
for the description of the Hoyle state. Taking into account
the cluster structure of $^{12}$C in  terms of three $\alpha$ particles provides an advantage to
reduce significantly the Hilbert space needed 
to describe $^{12}$C. 
In another model \cite{bijker,annphys}, claimed to be a cluster model,
a simple geometric structure
was assumed, namely a triangle with an $\alpha$ particle at each corner.
Though this model works quite well, it ignores the PEP, one of the most important principles
in nature. The motivation of our contribution is to compare the model 
in \cite{bijker}
to the SACM and test to what extend the PEP plays a role in $^{12}$C.
One attempt was already taken in \cite{arxiv} using the
multi-channel symmetry, which relates different clusterizations to
the same Hamiltonian. Here, we will discuss the $^{12}$C nucleus
as a three-$\alpha$ cluster system.

The paper is organized as follows: In section \ref{sec2} an introduction to the basic concepts of
the SACM with three clusters is presented. Also in section \ref{sec2} the model space is constructed, first for
the two-$\alpha$ subsystem $^{8}$Be and then the third $\alpha$ particle is added. It is shown
that the final model space can also be obtained considering a two-cluster system of $^8$Be+$\alpha$.
In section \ref{sec3} some geometrical considerations are presented, where
simple symmetry arguments and a geometrical mapping 
leads to a triangular structure of $^{12}$C in its ground state.
In Section \ref{sec4} 
a Hamiltonian is proposed, being a combination of a dominant $SU(3)$ part with one mixing term.
The spectrum is reproduced in the case of the $SU(3)$ limit
with the same quality as in \cite{bijker}.
With the mixing it is even better than in \cite{annphys}.
It will be shown that the mixing of $SU(3)$ irreps is crucial and 
that taking into account the PEP
is important.
It is also shown that
the cluster model permits $1^+$ states at low energy, which are forbidden
in the oblate symmetric top model \cite{bijker}. 

\section{The SACM and the construction of the model space for $^{12}$C}
\label{sec2}

For light nuclei, the $SU(3)$ symmetry is approximately realized. 
A useful basis is available
when the nucleus is divided into several clusters.
Within the SACM \cite{sacm1,sacm2}, 
each cluster is represented by an $SU(3)$-shell model irreducible representation
(irrep) $\left( \lambda_k, \mu_k\right)$, where $k$ refers to the number of the cluster.  The clusters are not excited because this would lead to
a double counting of shell model states, as will be illustrated further below.

For a three-cluster system, the model space in constructed in four steps:\\
1) Multiply the first {\it two cluster} irreps with the irrep of their relative motion:

\beqa
(\left(\lambda_1,\mu_1\right) \otimes \left(\lambda_2,\mu_2\right) \otimes \left(n_\rho , 0\right)
& = & 
\sum_{\lambda_{12},\mu_{12}} 
m_{\left(\lambda_{12},\mu_{12}\right)}\left(\lambda_{12},\mu_{12}\right)
~~~,
\label{eq-1}
\eeqa
where $m_{\left(\lambda_{12},\mu_{12}\right)}$ is the multiplicity of the corresponding irrep.
The $\rho$ denotes the relative motion between the first two $\alpha$-particles, whose relative vector
is $\vec{\rho}$ and $n_\rho$ is the number of relative oscillation quanta. Due to the
{\it Wildermuth condition} \cite{wilder} there is a lower limit for 
$n_{\rho 0}$. This number is
determined by the difference of the oscillation quanta in the parent nucleus
and
the sum of the oscillation quanta in each cluster. For example, in a two-$\alpha$ system, each
$\alpha$ particle carries zero quanta, i.e. the sum is zero. The united system $^{8}$Be has
four nucleons in the s-shell and four in the p-shell, thus, it carries 4 quanta. The difference
is therefore 4 quanta which is the value for $n_{\rho 0}$. Using a lower number would imply
that at least one nucleon has to be put into an already occupied state, which is forbidden
by the PEP. Thus, the {\it Wildermuth condition} is necessary in order 
to satisfy the PEP. 
A further restriction in the list can be applied for a symmetric two-cluster system \cite{sacm1}. 
For a two-$\alpha$-cluster system, $(-1)^{n_\rho}$ has to be positive, thus
only even $n_\rho$ are allowed. Because the $^8$Be has to be in its ground
state $SU(3)$ irrep, the $n_\rho$ is fixed to 4, but we also will discuss what happens when
the $^8$Be cluster is excited, in order to strengthen our argument,
concerning the double counting. 
\\
2) Take the result (\ref{eq-1}) and determine all the products
$\left(\lambda_3,\mu_3\right) \otimes$
$\left[\left(\lambda_{12},\mu_{12}\right)\right.$ $\otimes$ 
$\left.\left( n_\lambda , 0\right) \right]$. The $n_\lambda$
denotes the number of relative quanta of the third cluster to the center of mass of the first 
two clusters and $\vec{\lambda}$ is its relative vector.
The $n_\lambda$ is determined in the same manner as in 1); 
Assuming that $^8$Be is in its ground state $SU(3)$ irrep,
adding a further
$\alpha$ cluster, the two-$\alpha$ subsystem ($^{8}$Be) already carries 4 quanta, while the third
$\alpha$ particle carries zero quanta, the sum is 4. The united nucleus $^{12}$C carries 8 quanta,
having 8 nucleons in the p-shell. Thus, the Wildermuth condition is $n_{\lambda 0} = 4$.\\
3) The third cluster is combined with each $(\lambda_{12},\mu_{12})$ and the
result can be schematically written as a sum, namely

\beqa
\left(\lambda_3,\mu_3\right) \otimes
\left[\left(\lambda_{12},\mu_{12}\right) \otimes \left( n_\lambda , 0\right) \right] = 
\sum_{\left(\lambda , \mu \right)} m_{\left(\lambda , \mu \right)} \left(\lambda , \mu \right)
~~~,
\label{eq-2}
\eeqa
where $m_{\left(\lambda , \mu \right)}$ denotes the multiplicity of
$\left(\lambda , \mu \right)$.

The list of irreps in (\ref{eq-2}) with (\ref{eq-1})
still contains in general irreps which are
not allowed by the PEP. The SACM provides us with a method to eliminate the
non-allowed states:

\noindent
4) In each step, the shell model space for the corresponding sub-cluster system 
is constructed. For example, in the
two-$\alpha$ cluster subsystem, construct the shell model space of $^{8}$Be and determine
the overlap with the shell model space of the two-cluster system, which results in 
the SACM-model space of the 2-cluster subsystem.
When the remaining list is coupled with the third cluster and $\left(n_{\lambda},0\right)$,
the resulting list has to be matched with the shell model space of $^{12}$C.
Also to mention is that in each step the center of
mass has to be subtracted, which is particularly easy within the harmonic oscillator basis, for
details please consult \cite{yepez2012}.

\vskip 0.5cm

The steps 1) to 4) are now applied more specifically
to the three $\alpha$-cluster system:
In the two-$\alpha$ cluster sub-system we obtained for the
minimal number of relative oscillation quanta 4. The product of all irreps can be resumed by

\beqa
\left( 0,0\right) \otimes \left( 0,0 \right) \otimes \left(n_\rho = 4 ,0\right)
& = & \left(4 , 0 \right)
~~~.
\label{eq-3}
\eeqa

Suppose that the $^8$Be is excited to (6,0). It now carries 2 more quanta, which implies that the minimal number of $n_\lambda$ quanta so satisfy the PEP is 
reduced to 2. Calculating the overlap of the space obtained 
via (\ref{eq-2}) with the shell model leads to a list of allowed irreps 
{\it which is contained in the list} 
when the $^8$Be cluster is in the (4,0) irrep.
Thus, in order
to avoid a double counting the $^8$Be nucleus has to be in its ground 
state irrep. Exciting the $^8$Be
cluster to larger $n_\rho$ leads to the same observation.

When  the third $\alpha$ particle is added we obtain through the multiplication of $(0,0)$
again $(4,0)$. This irreps has to be multiplied with $(n_\lambda ,0)$,
with $n_\lambda \ge n_{\lambda 0}$, i.e. for $0\hbar\omega$ it is $(4,0)_{{\rm Be}}\otimes (4,0)$, for
$1\hbar\omega$ it is $(4,0)_{{\rm Be}} \otimes (5,0)$, for $2\hbar\omega$ it is
$(4,0)_{{\rm Be}} \otimes (6,0)$, etc. 

In Table \ref{tab1} the $SU(3)$ content of up to 
$6\hbar\omega$ excitations are listed, i.e. it represents the model 
space of the SACM
for $^{12}$C as a three $\alpha$ cluster system.

\begin{center}
\begin{table}[h!]
\centering
\begin{tabular}{|c|c|}
\hline\hline
$n\hbar\omega$ & $\left(\lambda , \mu\right)$ \\
\hline
0 & (0,4) \\
1 & (3,3)  \\
2 & (2,4), (4,3), (6,2)  \\
3 & (3,4), (5,3), (7,2), (9,1)  \\
4 & (4,4), (6,3), (8,2), (10,1), (12,0)  \\
5 & (5,4), (7,3), (13,0)  \\
6 & (6,4), (8,3), (10,2), (12,1)  \\
\hline
 \end{tabular}
\caption{
Model space of the $^{12}$C nucleus within the SACM for up to $6\hbar\omega$ excitations. 
Note, that for $n=2$, the (4,3) irrep contains a positive
parity, spin one state.
} 
\vspace{0.2cm}
\label{tab1}
\end{table}
\end{center}

Note, that the model space is microscopic, i.e., it observed the PEP!

From Table \ref{tab1} one can already infer the spin content of the spectrum,
using the rules of J.P. Elliott \cite{elliott}.
For even $n$, the states have positive parity and for odd $n$, the states have
negative parity. The lowest 
negative parity states are contained in the (3,3) irrep which has two
bands \cite{elliott}, one with $L=1^-$, $2^-$, $3^-$ and $4^-$ 
(it does not contain a $5^-$ state!) and the other band with
$3^-$, $4^-$, $5^-$ and $6^-$. Note, there is only one $5^-$ state at
low energy.
When the $SU(3)$-breaking term is switched on, these
states become mixed with higher lying $SU(3)$ irreps of the same parity.
As it will turn out further below, 
the Hoyle state does not come from a $2\hbar\omega$
excitation but from a $4\hbar\omega$ one. Note also, that there are
several $1^+$ states possible, which are forbidden in the oblate symmetric top
model of \cite{bijker}. For example, the (4,3) irrep at 2$\hbar\omega$
contains one $1^+$ state.

\section{Geometrical considerations}
\label{sec3}

In \cite{geom} a geometrical mapping of the SACM Hamiltonian was presented. 
The potential
is defined as the expectation value of this Hamiltonian with respect to a 
coherent state,
which takes into account the Wildermuth condition. The same structure can be used here, namely
a direct product of  a coherent state for the two-cluster subsystem and one for the $^{8}$Be+$\alpha$
system. The result gives a hint to the ground state configuration of a cluster system.

Due to the PEP two $\alpha$-clusters 
have a finite minimal distance, i.e., they cannot
overlap completely with a vanishing relative distance. 
The relation obtained \cite{geom} can be readily
extended to the three-cluster system of $^{12}$C and its is given by

\beqa
\rho_0 & = & \sqrt{\frac{\hbar n_{\rho 0}}{{\mu_{\alpha\alpha}} \omega}}
~~,~~
\lambda_0 ~ = ~ \sqrt{\frac{\hbar n_{\lambda 0}}{{\mu_{Be\alpha}} \omega}}
~~~,
\label{eq-4}
\eeqa
where $\mu_{\alpha\alpha}$ is the reduced mass of the two-$\alpha$ subsystem and
$\mu_{Be\alpha}$ the reduced mass of the $^{8}$Be+$\alpha$ cluster system.
Because the $\alpha$-clusters are indistinguishable (they are
symmetric under permutation) the distance $\rho_0$ between {\it any cluster}
has to be the same. This already indicates that the three $\alpha$-cluster
system has to be ordered in a triangle. When a $SU(3)$ mixing interaction is taken
into account, the $n_{\rho 0}$ and $n_{\lambda 0}$ values increase, thus,
the estimation given here yields lower limits, otherwise the consequences
about the geometric structure remain. 

We obtain $\rho_0 \approx \sqrt{\frac{80}{\hbar\omega}}$~fm and 
$\lambda_0 \approx 2\sqrt{\frac{15}{\hbar\omega}}$~fm, which is in a very nice
agreement to the microscopic cluster model of \cite{neef}. 
We took into account that $\mu_{\alpha\alpha}= \frac{m_\alpha}{2}$ and 
$\mu_{Be\alpha}=\frac{2}{3}m_\alpha$ (neglecting energy binding effects), 
with $m_\alpha$ being the mass of the $\alpha$ particle,
we obtain for the ratio of the distances

\beqa
\frac{\lambda_0}{\rho_0} & = & \frac{\sqrt{3}}{2}
~~~,
\label{eq-5}
\eeqa
which simply confirms the triangular structure (Pythagoras!)- 

This shows that with a few elementary considerations, without recurring
to ad hoc assumptions and/or complicated calculations, the ground state 
geometrical configuration of $^{12}$C is obtained!

\section{The Hamiltonian and results}
\label{sec4}

As a Hamiltonian we propose a combination of a pure $SU(3)$-part
and a symmetry breaking term, which is a generator of $O(4)$. The
Hamiltonian is also chosen to have for the $SU(3)$ limit
the same number of parameter used
in \cite{bijker}, where the cut-off N is used as one of the parameters.
The choice does not exclude the use of a more  general Hamiltonian.
As in \cite{bijker} the total number of quanta is 
$N=n_\rho + n_\lambda + n_\sigma$, where the $\sigma$-bosons are
introduced as a trick to obtain a cut-off. 

The model Hamiltonian proposed is

\beqa
{\bd H} & = & \hbar\omega {\bd n}_\pi
-\chi  {\bd {\cal C}}_2(\lambda , \mu ) 
+ t_2({\bd {\cal C}}_2(\lambda , \mu ))^2 
+ t {\bd {\cal C}}_3(\lambda , \mu ) 
\nonumber \\
&& + \left(a+a_L(-1)^L + a_{Lnp} \Delta {\bd n}_\pi \right) {\bd L}^2 +b{\bd K}^2
\nonumber \\
&& +b_1 \left[\left({\bd\sigma}^\dagger \right)^2 
- \left( {\bd \pi}^\dagger \cdot {\bd \pi}^\dagger \right) \right]
\cdot \left[h.c.\right]
~~~.
\label{eq-6}
\eeqa
The  first term is just the harmonic oscillator field and the $\hbar\omega$ 
is fixed 
via $45 \times A^{-1/3} - 25 \times A^{-2/3}$ \cite{hw}, where its value is 14.89
for $^{12}$C. The second term is related to
the quadrupole quadrupole interaction \cite{elliott}, with ${\bd {\cal C}}_2(\lambda , \mu )$
being the second order Casimir operator. 
The third and fourth terms allow for corrections in the relative ordering of $SU(3)$ irreps,
where ${\bd {\cal C}}_3$ is the third order Casimir operator.
The fifth term allows
to describe changes in the moment of inertia
for states with higher shell excitations and when the spin parity changes. 
The last term in the second line
lifts the degeneracy in angular 
momentum for states within the same $SU(3)$ irrep. 
Up to here, the Hamiltonian is within the $SU(3)$ limit
and permits analytic results, substituting the operators by their 
corresponding eigenvalues.
In the last line, the term mixes $SU(3)$ irreps, it is
a generator of a $O(4)$ group. 
The pure $SU(3)$ part has 7 free parameters, the same as in \cite{bijker} (including $N$), and will
be compared to the oblate top model. For allowing the mixing a further barameter ($b_1$) is added,
i.e., in the final calculations there are 8
parameters.

We will apply 3 different sample calculations: i) A pure $SU(3)$ Hamiltonian,
which permits the best adjustment to the spectrum, also obtained in \cite{bijker}.
ii) The $B(E2; 0_2^+ \rightarrow 2_1^+)$ is adjusted to about a tenth of
the experimental value, which is 8~Wu \cite{exp}, as done in \cite{annphys}. 
iii) the transition value is adjusted to the experimental value.

The $B(E2)$ value mentioned is essential,
because it signals an important mixture 
between $SU(3)$-bands. It will also make it difficult to adjust the 
spectrum and grouping states into bands, 
which is easier when a simple algebraic model is used.

The use of $N$ as a parameter in \cite{bijker,annphys} is questionable,
because the $N$ is a cut-off parameter.
In the oblate top model the $N$ is related to the 
anharmonicities of the potential. This
interpretation is often used and due to a misleading understanding of the
geometrical mapping, as we already investigated in 
\cite{huitz-2006,phase-I,phase-II}. 

As the quadrupole transition operator we use the one given in \cite{octavio},
which is a symplectic generator, including connections to multiple $2\hbar\omega$
shell excitations. In an algebraic model it is customary to use the algebraic part
of this operator, which does not connect shells, which is valid when inter-shell
excitations are not considered.
The physical quadrupole operator \cite{octavio,sympl} is given by

\beqa
{\bd Q}^{phys}_{2m} & = & Q^a_{2m} + \frac{\sqrt{6}}{2}
\left( {\bd B}^\dagger_{2m} + {\bd B}_{2m} \right)
\nonumber \\
{\bd B}^\dagger_{2m} & = & \left({\bd \pi}^\dagger \cdot {\bd \pi}^\dagger
\right) 
~,~
{\bd B}_{2m} ~=~ \left({\bd \pi} \cdot {\bd \pi} \right)
~~~.
\label{Qphys}
\eeqa
The ${\bd B}^\dagger_{2m}$ operator transforms as a (2,0) $SU(3)$ irrep,
while ${\bd B}_{2m}$ as its conjugate.

As we will see, in order to get a large transition value between
states in different bands, a strong mixing of the $SU(3)$ basis
is required, which will lead to a less favorable agreement in the spectrum.
This is a characteristic feature when simple models are used, as we do here too.
In order to obtain a better agreement, more interaction terms have to be included,
which will increase the number of parameters, from which we will refrain here for the sake
of comparison. We also want to show that sometimes simple models are not enough!

\begin{table}
\begin{center}
\begin{tabular}{|c|c|c|c|c|}
\hline
$B(EL;J^\pi_i \rightarrow J^\pi_f$ [WU] & EXP. & $SU(3)$ & case-1 & case-8 \\
\hline
$B(E2;2^+_1 \rightarrow 0^+_1$ & 4.65 & 4.65 & 5.18 & 3.41 \\ 
$B(E2;0^+_2 \rightarrow 2^+_1$ & 8. & 0.0 & 0.97 & 8.33 \\ 
$B(E3;3^-_1 \rightarrow 0^+_1$ & 12. & 6.32 & 12.23 & 8.64 \\ 
\hline
 \end{tabular}
\end{center}
\caption{\label{table2}
List of $B(EL)$-transition values, measured and 
obtained in three different model calculations:
In the first column information is listed on the type of the electro-magnetic
transition, the second column lists the corresponding experimental value,
the third column assumes exact $SU(3)$ symmetry and in the last two columns
the experimental value of $B(E2;2^+_2 \rightarrow 0^+_1)$ is adjusted to
different values (case-1: 1WU, case-8: 8WU), where an increased value corresponds to a larger
$SU(3)$ mixing. The E3- transition values are not adjusted.
}
\end{table}

\begin{table}
\begin{center}
\begin{tabular}{|c|c|c|c|}
\hline
parameter [MeV] & $SU(3)$ & case-1 & case-8  \\
\hline
$\chi$ &  -0.190031 & -0.759075 & -0.988014 \\
$t_2$ & 0.00150477 & -0.000490236  & -0.00275029 \\ 
$t$ & -0.0124853 & 0.0160426 &  0.0363740 \\ 
$a$ & 0.669745 & 0.459184 & 0.592743 \\ 
$a_L$ & -0.0269914 & 0.00639953 & -0.00740967 \\ 
$a_{Lnp}$ &  -0.0854681 & -0.00178681 & -0.00408504 \\ 
$b$ & -0.776579 & -0.575894 & -0.594000 \\ 
$b_1$ & 0.0 & -1.429287  & -0.510012 \\ 
\hline
 \end{tabular}
\end{center}
\caption{\label{table3}
List of the parameter values used. The first column lists the parameter
symbols, the second their numerical value for the $SU(3)$ limit,
the third for the case when the transition value $B(E2; 0_2^+ \rightarrow 2_1^+)$
is fitted to 1 (case-1) and the fourth when it is fitted to 8 (case-8). 
}
\end{table}

\begin{table}
\begin{center}
\begin{tabular}{|c|c|c|c|c|c|c|}
\hline
$L^\pi_i$ & $0_1^+$ & $2_1^+$ & $4_2^+$ & $0_2^+$ & $2_2^+$ & $4_1^+$\\ 
\hline
$n_\pi = 4$ :(0,4) & 62 & 61 & 22 & 4 & 1 & 36 \\
$n_\pi = 6$ :(2,4) & 29 & 28 & 5 & 0 & 0 & 25 \\ 
$n_\pi = 6$ :(4,3) & 0 & 0 & 15 & 0 & 12 & 9 \\ 
$n_\pi = 8$ :(4,4) & 5 & 6 & 3 & 1 & 0 & 0 \\ 
$n_\pi = 8$ :(6,2) & 2 & 3 & 22 & 18  & 43 & 4 \\
$n_\pi = 8$ :(6,3) & 0 & 0 & 13 & 0 & 9 & 5 \\
$n_\pi = 8$ :(8,2) & 1 & 1 & 13 & 17  & 26 & 2 \\
$n_\pi = 8$ :(12,0) & 0 & 0 & 0 & 58  & 2 & 0 \\
$n_\pi = 10$ :(6,4) & 0 & 0 & 0 & 0 & 0 & 2 \\ 
$n_\pi = 10$ :(8,3) & 0 & 0 & 0 & 0 & 1 & 1 \\ 
$n_\pi = 10$ :(10,2) & 0 & 0 & 2 & 2 & 3 & 0 \\ 
\hline
 \end{tabular}
\end{center}
\caption{\label{table3.1}
$SU(3)$ content of some low lying states with positive parity, given
in percent, for the case of $B(E2; 0_2^+ \rightarrow 2_1^+)=1~WU$. 
The numbers are only approximate and not all irreps are shown.
The first group refers to the ground state band and the second one to the
Hoyle state.
Clearly seen is the similar structure of the $0_1^+$ and $2_1^+$ state,
some similarity of the $0_2^+$ to the $2_2^+$ state and the similarity
vanishes for the $4^+$ states. 
}
\end{table}

\begin{table}
\begin{center}
\begin{tabular}{|c|c|c|c|c|c|}
\hline
$L^\pi_i$ & $1_1^-$ & $2_1^-$ & $3_1^-$ & $4_1^-$ & $5_1^-$ \\ 
\hline
$n_\pi = 5$ :(3,3) & 54 & 52 & 59 & 54 & 41 \\
$n_\pi = 7$ :(3,4) & 5 & 7 & 4 & 10 & 10 \\ 
$n_\pi = 7$ :(5,3) & 31 & 30 & 30 & 26 & 30  \\ 
$n_\pi = 7$ :(7,2) & 0 & 0 & 1 & 1 & 4  \\ 
$n_\pi = 7$ :(9,1) & 3 & 3 & 0 & 1 & 1 \\ 
$n_\pi = 9$ :(5,4) & 1 & 3 & 0 & 4 & 0  \\ 
$n_\pi = 9$ :(7,3) & 5 & 5 & 5 & 4 & 7 \\
\hline
 \end{tabular}
\end{center}
\caption{\label{table3.2}
$SU(3)$ content of some low lying states with negative parity, given as 
in percent, for the case of $B(E2; 0_2^+ \rightarrow 2_1^+)=1~WU$. 
The numbers are only approximate and not all irreps are shown.
The first group refers to the ground state band and the second one to the
Hoyle state.
Clearly seen is the similar structure of the $1_1^-$ and $2_1^-$ state,
the same for the $3_1^-$ to the $5_1^-$. 
}
\end{table}

\begin{table}
\begin{center}
\begin{tabular}{|c|c|c|c|c|c|c|}
\hline
$L^\pi_i$ & $0_1^+$ & $2_1^+$ & $4_2^+$ & $0_2^+$ & $2_2^+$ & $4_1^+$\\ 
\hline
$n_\pi = 4$ :(0,4) & 82 & 82 & 48 & 8 & 3 & 31 \\
$n_\pi = 6$ :(2,4) & 16 & 16 & 4 & 22 & 41 & 16 \\ 
$n_\pi = 6$ :(4,3) & 0 & 0 & 0 & 0 & 0 & 0 \\ 
$n_\pi = 8$ :(4,4) & 1 & 2 & 0 & 22 & 28 & 29 \\ 
$n_\pi = 8$ :(6,2) & 0 & 0 & 0 & 5  & 6 & 0 \\
$n_\pi = 8$ :(6,3) & 0 & 0 & 0 & 0 & 0 & 0 \\
$n_\pi = 8$ :(8,2) & 0 & 0 & 0 & 2  & 6 & 0 \\
$n_\pi = 8$ :(12,0) & 0 & 0 & 0 & 33  & 6 & 0 \\
$n_\pi = 10$ :(6,4) & 0 & 0 & 39 & 7 & 9 & 19 \\ 
$n_\pi = 10$ :(10,2) & 0 & 0 & 0 & 0 & 0 & 0 \\ 
\hline
 \end{tabular}
\end{center}
\caption{\label{table3.3}
$SU(3)$ content of some low lying states with positive parity, given as 
in percent, for the case of $B(E2; 0_2^+ \rightarrow 2_1^+)=8~WU$.
The numbers are only approximate and not all irreps are shown.
The first group refers to the ground state band and the second one to the
Hoyle state.
Clearly seen is the similar structure of the $0_1^+$ and $2_1^+$ state,
some similarity of the $0_2^+$ to the $2_2^+$ state and the similarity
vanishes for the $4^+$ states. 
}
\end{table}

\begin{table}
\begin{center}
\begin{tabular}{|c|c|c|c|c|c|}
\hline
$L^\pi_i$ & $1_1^-$ & $2_1^-$ & $3_1^-$ & $4_1^-$ & $5_1^-$ \\ 
\hline
$n_\pi = 5$ :(3,3) & 78 & 74 & 82 & 68 & 14 \\
$n_\pi = 7$ :(3,4) & 5 & 7 & 2 & 10 & 11 \\ 
$n_\pi = 7$ :(5,3) & 15 & 14 & 10 & 12 & 11  \\ 
$n_\pi = 7$ :(7,2) & 0 & 0 & 0 & 0 & 0  \\ 
$n_\pi = 7$ :(9,1) & 0 & 0 & 0 & 0 & 0 \\ 
$n_\pi = 9$ :(5,4) & 1 & 4 & 0 & 9 & 59  \\ 
$n_\pi = 9$ :(7,3) & 1 & 1 & 1 & 1 & 4 \\
\hline
 \end{tabular}
\end{center}
\caption{\label{table3.4}
$SU(3)$ content of some low lying states with negative parity, given as 
in percent, for the case of $B(E2; 0_2^+ \rightarrow 2_1^+)=8~WU$.
The numbers are only approximate and not all irreps are shown.
The first group refers to the ground state band and the second one to the
Hoyle state.
Clearly seen is the similar structure of the $1_1^-$ and $2_1^-$ state,
the same for the $3_1^-$ to the $4_1^-$. However, the similarity end for
the $5_1+$ state state.
}
\end{table}

\begin{table}
\begin{center}
\begin{tabular}{|c|c|c|c|c|c|c|c|c|c|c|c|c|}
\hline
state & $0_1^+$ & $2_1^+$ & $4_1^+$ & $0_2^+$ \\
\hline
spec. fact. & 0.0866 & 0.0218 & 0.00594 & 0.0417 \\
\hline
state & $2_2^+$ & $4_2^+$ & $1_1^-$ & $2_1^-$ \\
\hline
spec. fact. & 0.0191 & 0.00662 & 0.0160 & 0.0481 \\
\hline
state & $3_1^-$ & $4_1^-$ & $5_1^-$ & $0_3^+$ \\
\hline
spec. fact. & 0.000818 & 0.000909 & 0.00504 & 0.0291 \\
\hline
state & $1_1^+$ & $1_2^+$ & $1_2^-$ & $2_2^-$ \\
\hline
spec. fact. & 0.0449 & 0.0133 & 0.0024 & 0.053 \\
\hline
 \end{tabular}
\end{center}
\caption{\label{table4}
Spectroscopic factors for $^8$Be+$\alpha$ $\rightarrow$ $^{12}$C. The first,
third, fifth and seventh
rows indicate the state in $^{12}$C and the second, fourth, sixth and eighth 
rows list the
spectroscopic factor values. 
The states are ordered into bands, such that the energy
of the $4_1^+$ state has a larger energy than $4_2^+$. 
}
\end{table}

\begin{figure}
\begin{center}
\rotatebox{270}{\resizebox{120pt}{120pt}{\includegraphics[width=0.23\textwidth]{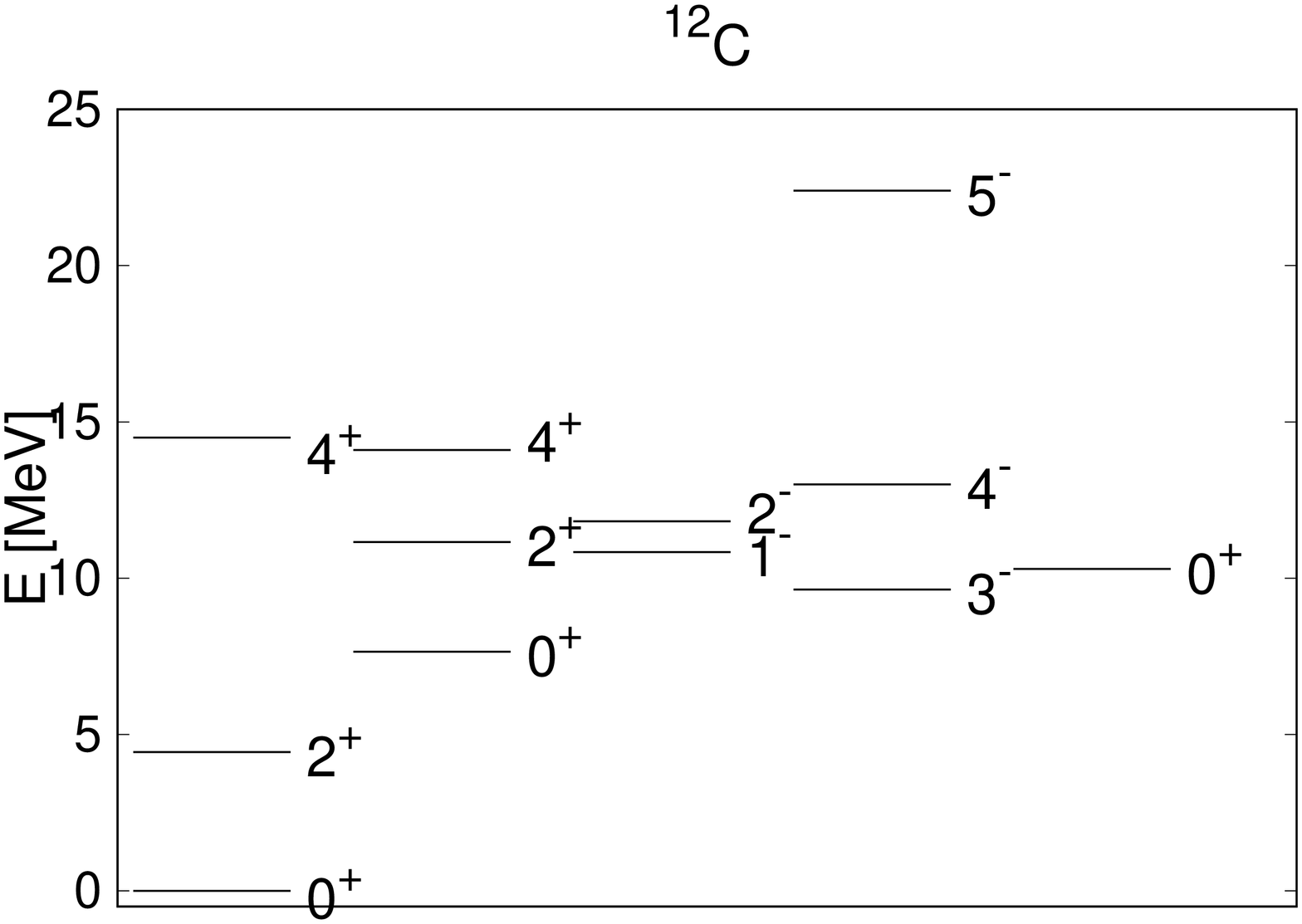}}} 
\rotatebox{270}{\resizebox{120pt}{120pt}{\includegraphics[width=0.23\textwidth]{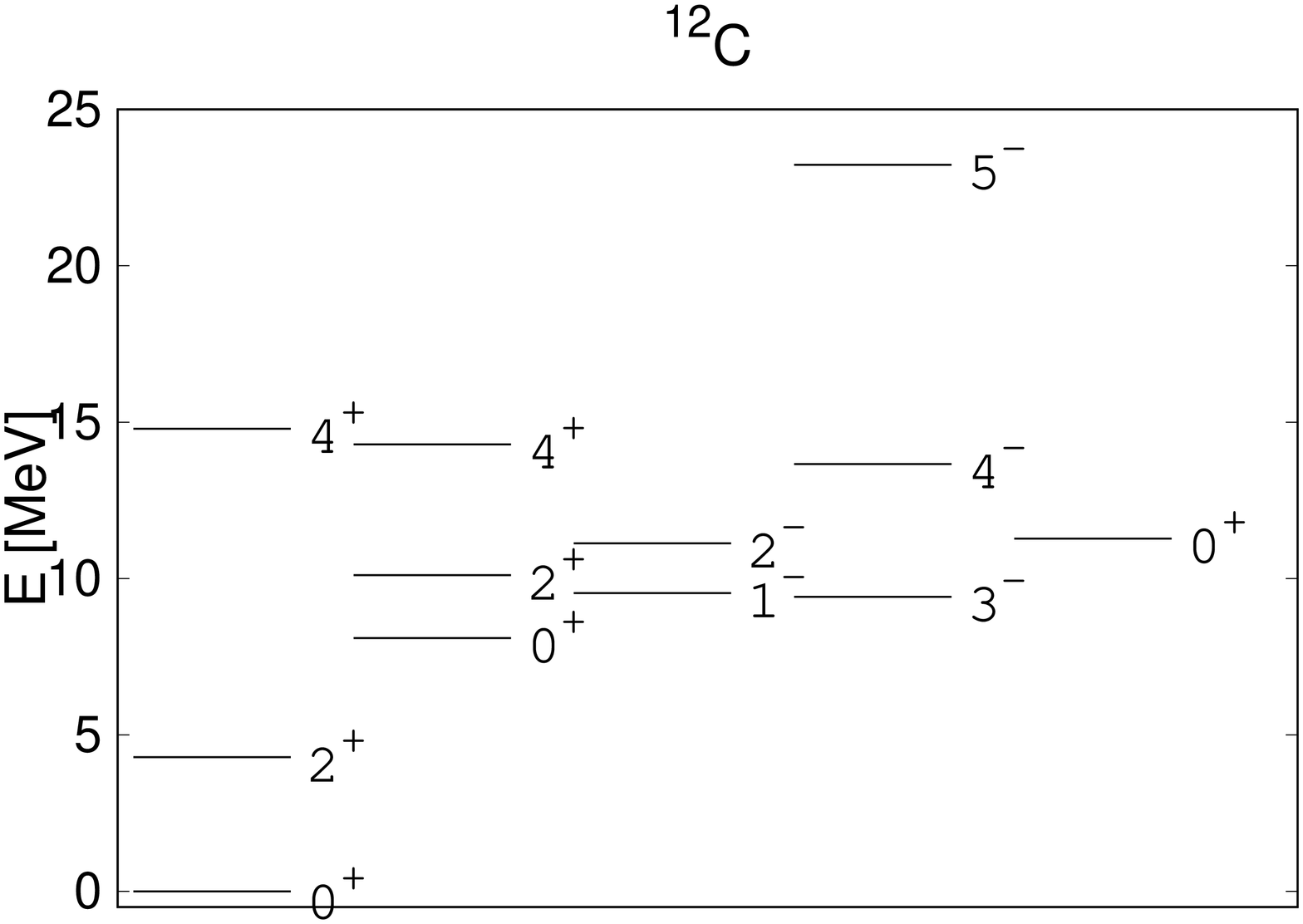}}} \\
\rotatebox{270}{\resizebox{120pt}{120pt}{\includegraphics[width=0.23\textwidth]{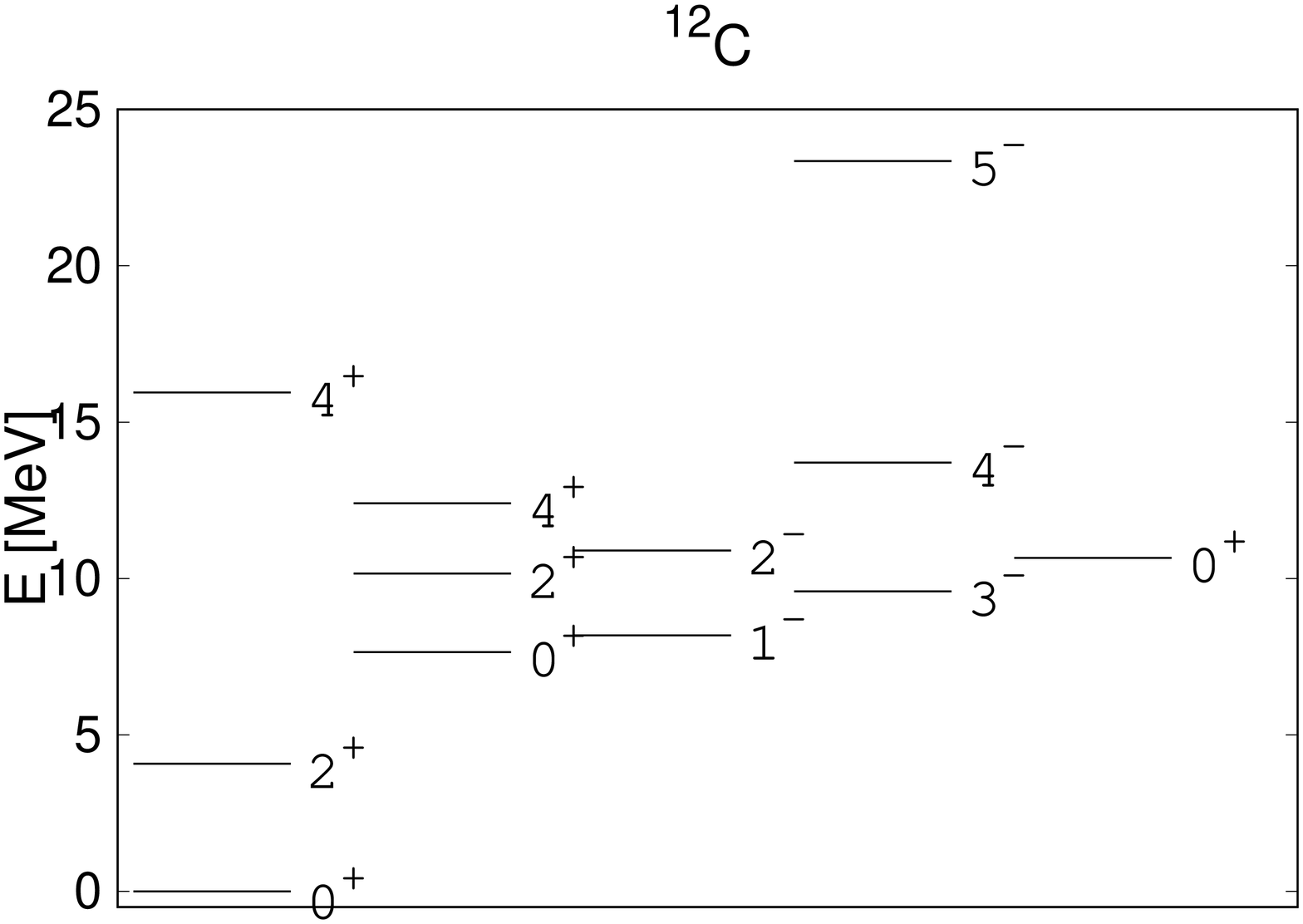}}} 
\rotatebox{270}{\resizebox{120pt}{120pt}{\includegraphics[width=0.23\textwidth]{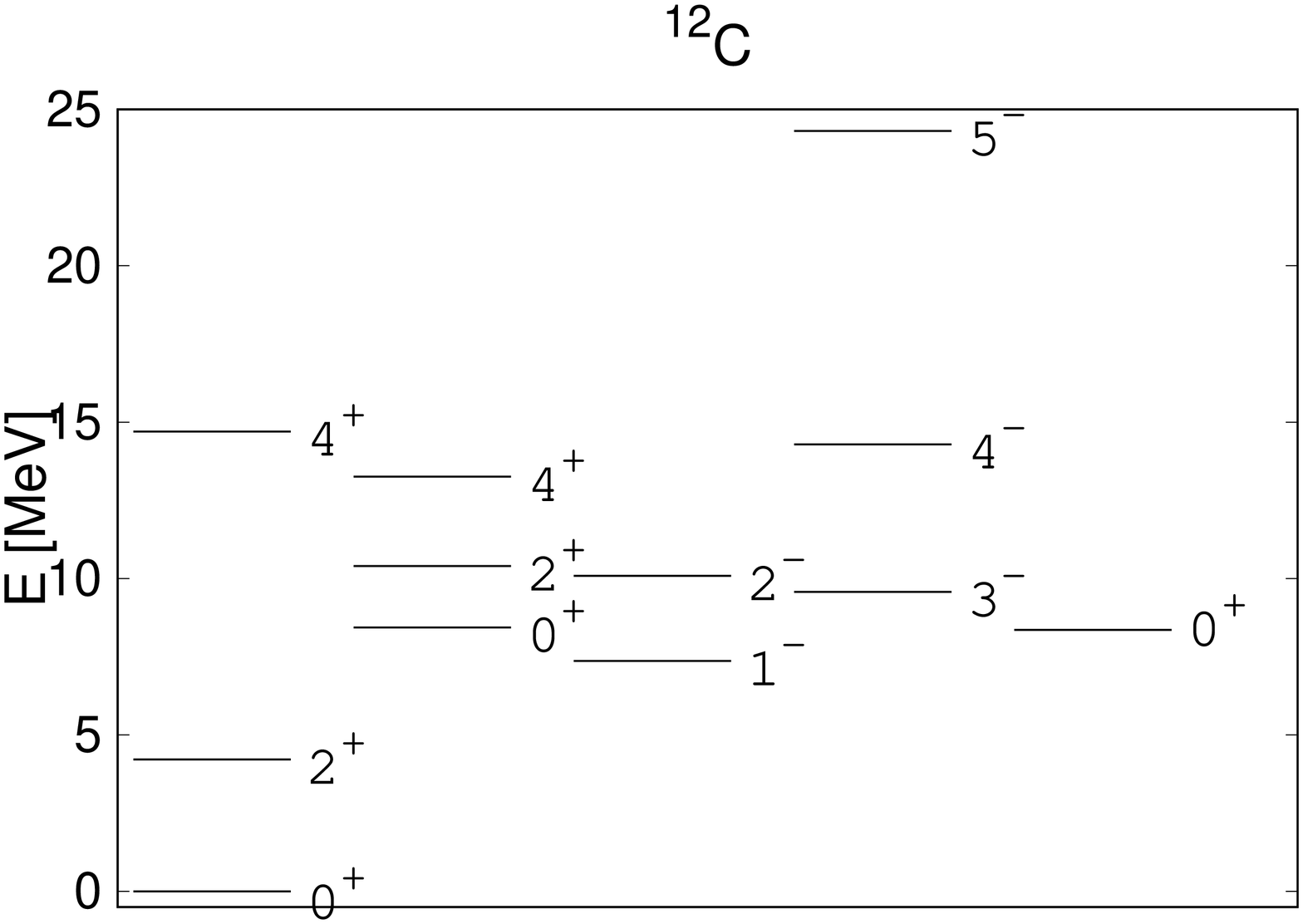}}} 
\caption{\label{fig1}
Spectrum of $^{12}$C. 
In the upper row, the left figure is the experimental spectrum and the right
figure depicts the result for pure $SU(3)$. In the second row, the left
figure depicts the result 
when the $B(E2;0_2^+ \rightarrow 2_1^+)$ transition value is
adjusted to 1~WU and the right figure when this value is 
adjusted to the experimental one, namely 8~WU.
}
\end{center}
\end{figure}

\begin{figure}
\begin{center}
\rotatebox{270}{\resizebox{120pt}{120pt}{\includegraphics[width=0.23\textwidth]{12C-energy-E2.eps}}} 
\rotatebox{270}{\resizebox{120pt}{120pt}{\includegraphics[width=0.23\textwidth]{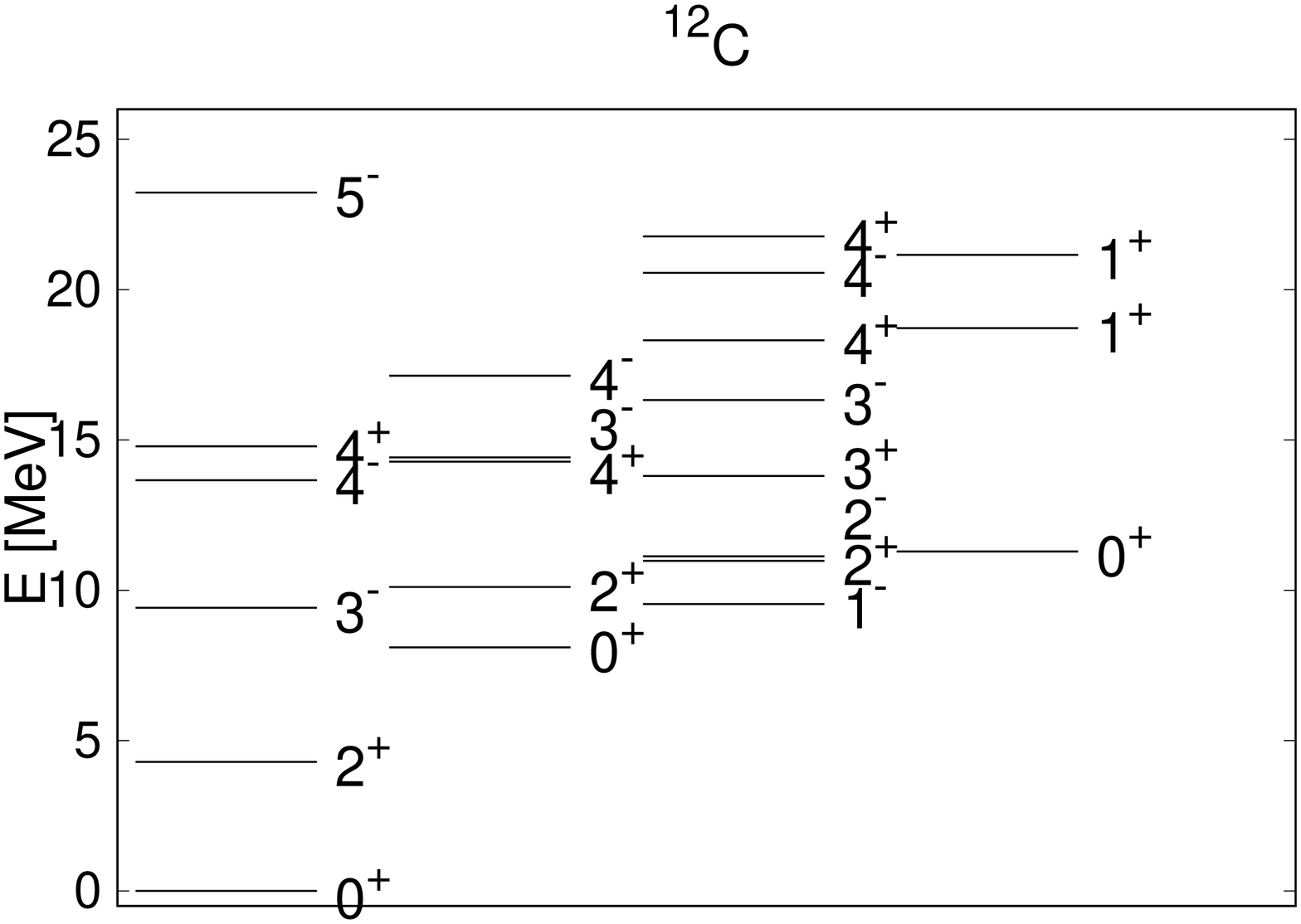}}} \\
\rotatebox{270}{\resizebox{120pt}{120pt}{\includegraphics[width=0.23\textwidth]{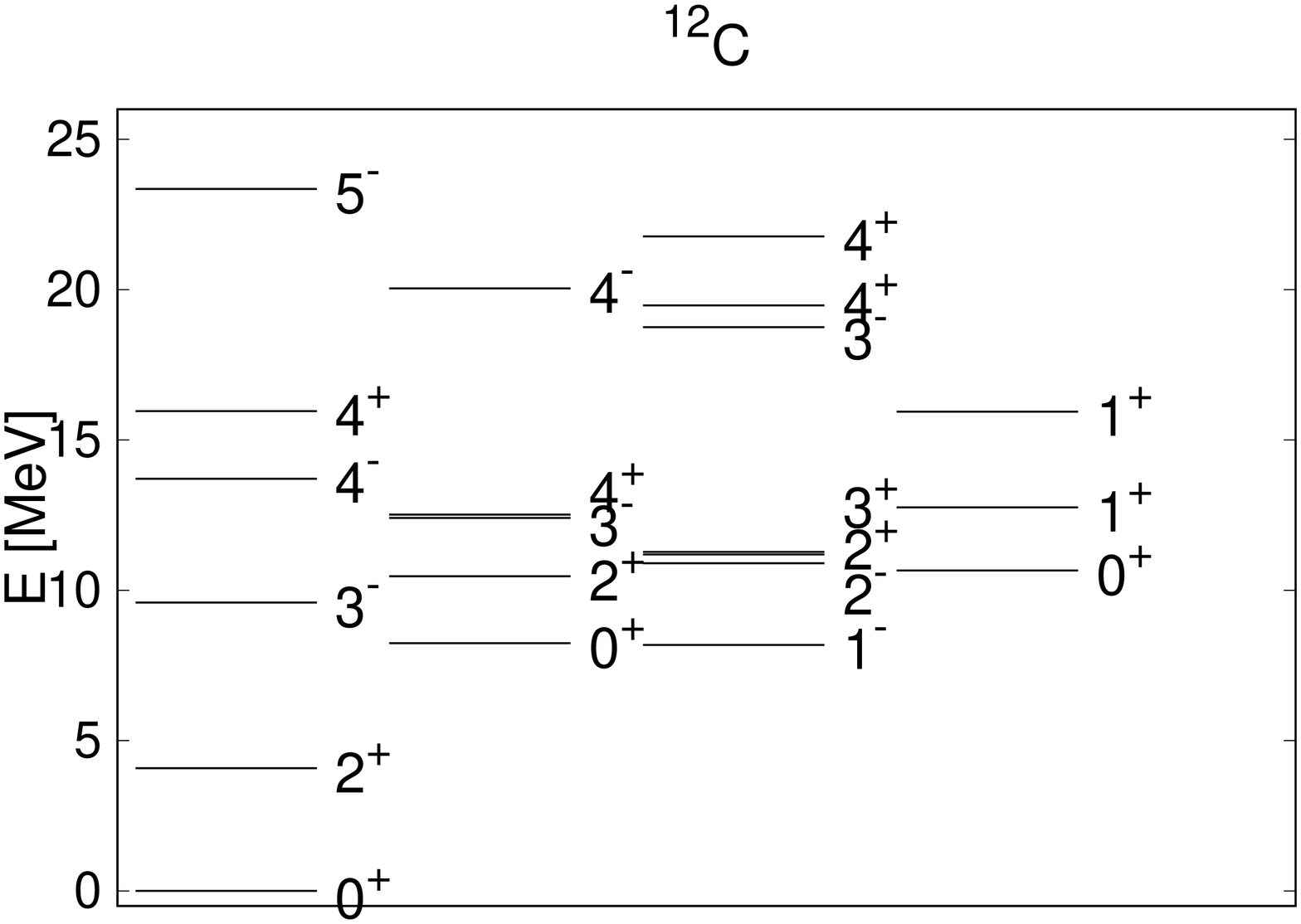}}}
\rotatebox{270}{\resizebox{120pt}{120pt}{\includegraphics[width=0.23\textwidth]{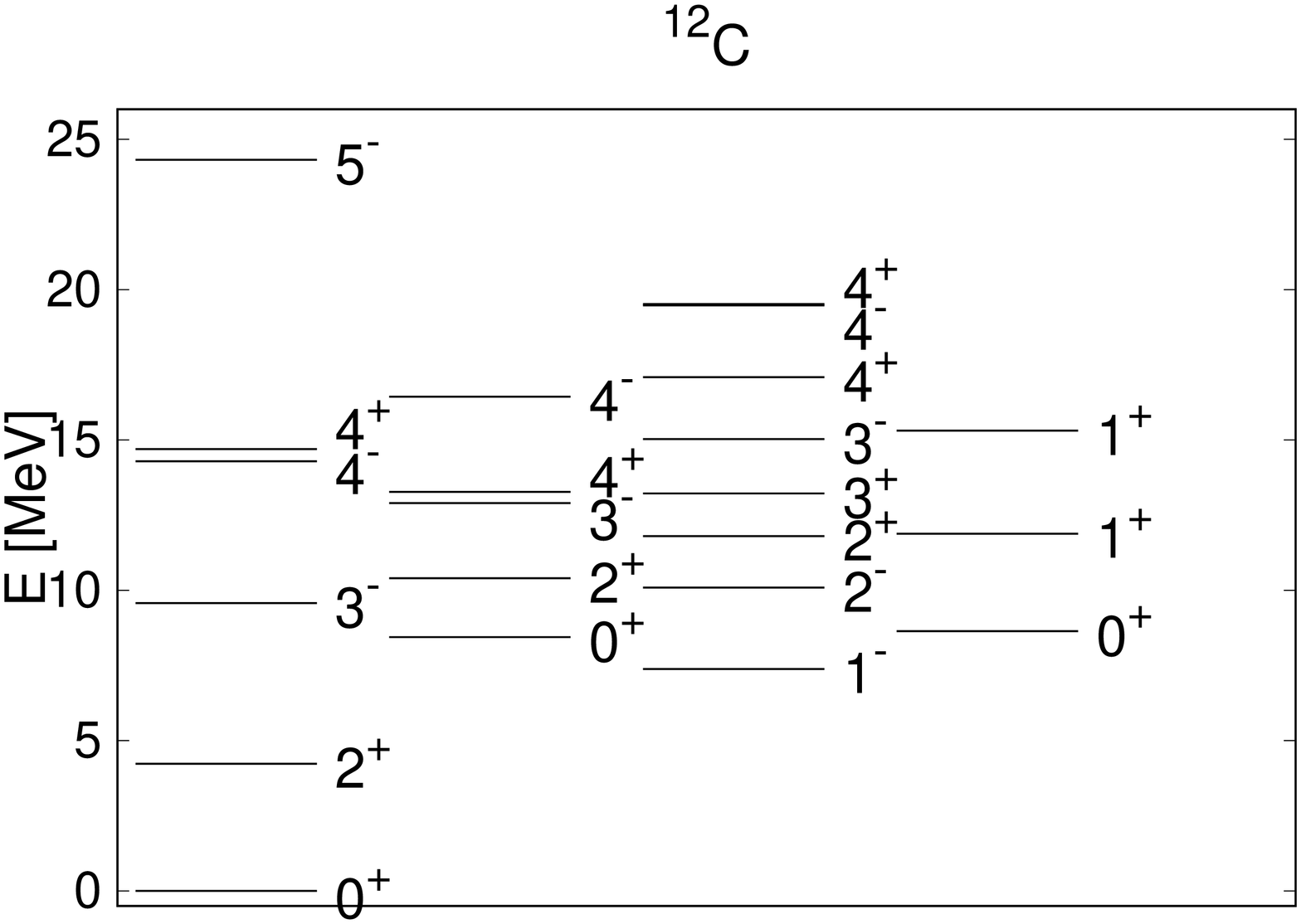}}}
\caption{\label{fig3}
Spectrum of $^{12}$C up to 25~MeV, using the same grouping into bands as in
\cite{bijker}.
The $B(E2;0_2^+ \rightarrow 2_1^+)$ transition value is
adjusted to the experimental one, namely 8~WU.
}
\end{center}
\end{figure}

\begin{figure}
\begin{center}
\rotatebox{270}{\resizebox{200pt}{200pt}{\includegraphics[width=0.23\textwidth]{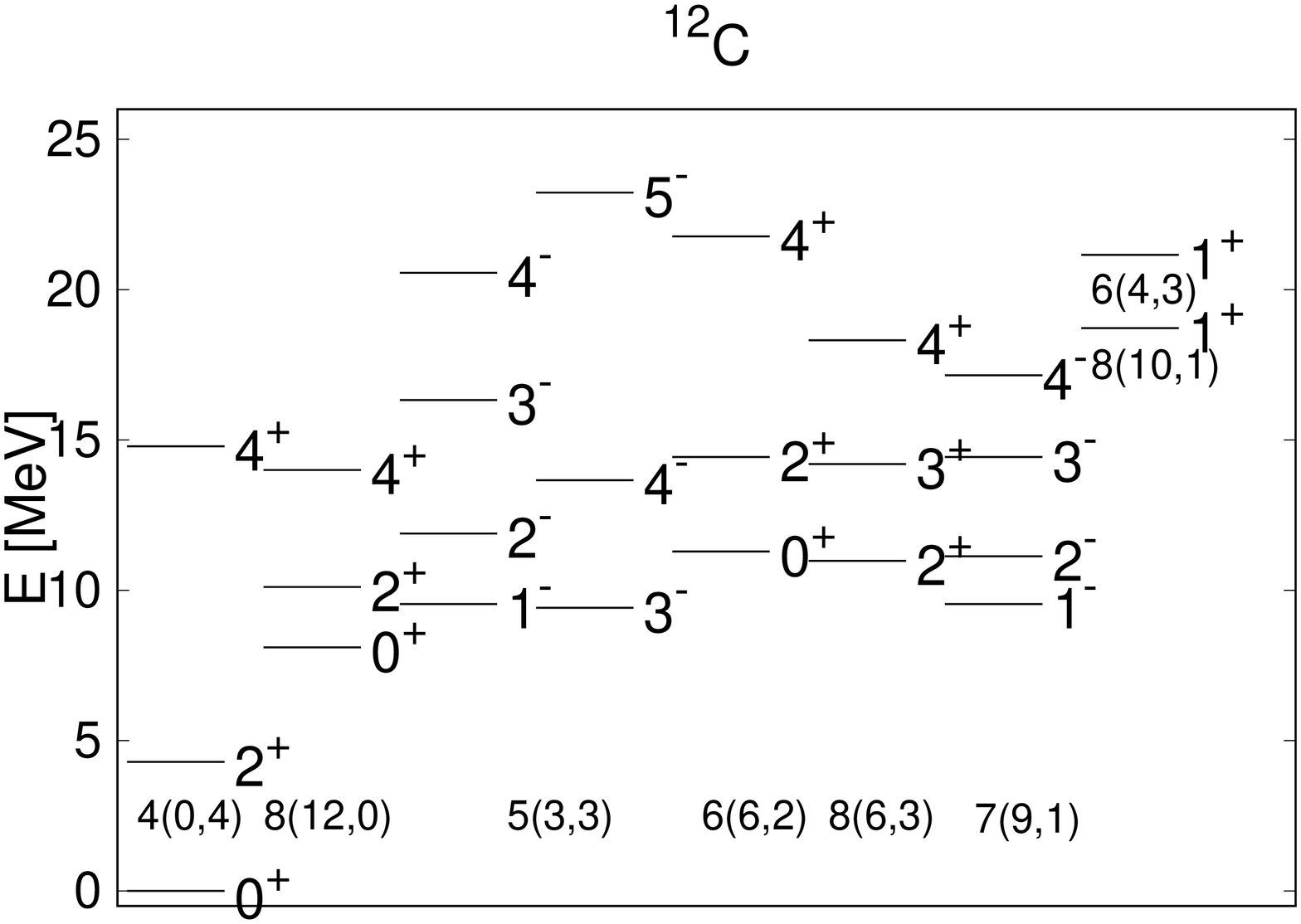}}} 
\caption{\label{fig4}
Spectrum of $^{12}$C up to 25~MeV,  grouping the states into bands according
to their $SU(3)$ irreps, which is listed below the bands.
Note the additional $1^-$-band head. The lowest $1_+$ state is the band head
of $n_\pi=8$~(10,1) and the second one of $n_\pi=6$~(4,3)~$K=1$.
}
\end{center}
\end{figure}

Though, for light nuclei the $SU(3)$ symmetry is well realized, there are 
strong indications in the electro-magnetic transitions that the $SU(3)$ irreps
mix significantly, due to the large deformation of $^{12}$C. 
In the $SU(3)$ limit, the second excited $0^+$ and $2^+$ states
belong to the irrep (12,0), a 4$\hbar\omega$ excitation, the first $2^+$
to (0,4) and the third $0^+$-band head to the (6,2) irrep at two excitation quanta,
all in agreement with the no-core shell model calculation in \cite{draayer-hoyle}. 
With increasing
$B(E2;0_2^+ \rightarrow 2_1^+)$ the mixing of $SU(3)$ gets larger.
We adjusted the spectrum and electro-magnetic transitions, assuming the three
different values for $B(E2;0_2^+ \rightarrow 2_1^+)$, discussed further above.
These three cases are listed in Table \ref{table2} and compared to the
experimental values, all in Weisskopf units. 
The E3-transition value indicates a large octupole component and it is reproduced
in all cases, without fitting it. This already shows that within the shell
model the deformation of the $^{12}$C nucleus is well described and does not
represent a particular success of any other model. 

In Figure \ref{fig1} the results are resumed, depicting only the up to 
now confirmed states.
In the $SU(3)$ limit the agreement to experiment is the best, as also
in \cite{bijker}, where the $B(E2;0_2^+ \rightarrow 2_1^+)$ is not listed.
In the left figure in the second row in Figure \ref{fig1} 
the experimental $B(E2;0_2^+ \rightarrow 2_1^+)$ is assumed to be 
1., which increases the mixing and as a consequence the agreement to
experiment is less favorable. With respect to the position of states and
the splitting within bands, the result is even better than in \cite{annphys},
where the theoretical $B(E2;0_2^+ \rightarrow 2_1^+)$ 
is listed as a tenth of the
experimental value. In the right figure in the second row of Figure \ref{fig1} 
the same $B(E2)$ value is adjusted to
8 \cite{exp}. Weisskopf units and consequently the mixing increases significantly
which results in a less good agreement for the spectrum. 

The fitting routine has the possibility to define a weight for a 
particular transition (also for the energy values). We increased the
weight for $B(E2;0_2^+ \rightarrow 2_1^+)$ step by step, starting
from zero. At first, the transition value stays very small and the spectrum has
a very good agreement to experiment, until at one point where the transition
value jumps to the large value which one tries to reproduce and at the same
time the agreement in the spectrum also jumps to be less favorable, which shows
that the particular transition makes it difficult to adjust the spectrum.
More interactions are needed for a good agreement, or a more involved model as the
no-core shell model calculations in \cite{draayer-hoyle}.

In Figure \ref{fig3} different spectra are shown, ordered in the same
manner as in \cite{bijker}, though we do not agree with this ordering
(see further below). We do
it simply for comparison. In the first row left panel the experimental result is 
repeated, followed by the $SU(3)$ result. In the second row the left panel 
refers
to the case when the transition value is adjusted to 1~WU, 
while the right panel
corresponds to 8~Wu. 
Only the $SU(3)$ case can be compared to the oblate top model.
An important feature are the spin-doublets in the ground state band, the 
Hoyle-band and the $1^-$-band (again, we use the association used in \cite{bijker}).
The $4^+-4^-$ doublets in the ground state
band and in the Hoyle-band are also obtained in the $SU(3)$ limit. The doublet
structure in the $1^-$-band, however, is only partially reproduced. 

One of the most important
feature is the appearance of $1^+$ cluster states, forbidden in the oblate top model. In Figure \ref{fig4} the spectrum, ordered according to the SACM into 
bands within the $SU(3)$ limit, is shown, including more states not
shown in the other figures.
A distinct difference is that in the $SU(3)$ limit there is no $5^-$ state in the Hoyle-band. Within the SACM a {\it second} $1^-$-band appears, nearly degenerate
with the first $1^-$-band. While the first $1^-$-band belongs to the
$n_\pi = 5$ (3,3) ($\Delta n_\pi = 1$) irrep of $SU(3)$, the second one comes from
$n_\pi = 7$ (9,1) ($\Delta n_\pi = 3$), which is significantly different to teh first 
$1^-$-band. 
Considering the mixing, the two
$1^-$-bands are not exactly degenerate but near to each other.

In general, the spectrum supports the triangular structure proposed in \cite{annphys,bijker},
though the signatures (spin-doublets, the missing $5^-$ state and the appearance of
$1^+$ cluster states) in the spectrum are not the same, once the PEP
is taken into account.

When the mixing is included, the spectrum becomes less and less similar to the experimental one,
though for the case when the transition value is adjusted to 1~WU, the agreement is still
acceptable to the experiment. The problems are related to the moment of inertia in the Hoyle
and the $1^-$-band. In the first one it is too large
while in the second one it is too low, though they are in much better agreement to
experiment than in \cite{annphys}. Also the position of the band heads get increasingly more
difficult to adjust. Some of the spin-doublets remain, however it is difficult to judge the 
case with a maximal mixing. The main indication is that the spin-doublets at larger energy will
be diluted. Again the missing $5^-$ state, the appearance of $1^+$ cluster states
and an additional $1^-$-band 
show the main difference to \cite{bijker}.

With respect to the association of bands, we do not agree to the interpretation
given in \cite{bijker}. As already mentioned, in the $SU(3)$ limit the content
of the positive parity states are not the same as for the negative parity states. Thus, they
cannot be ordered into the same band, a consequence of the PEP.
The content of the states in terms of the $SU(3)$ basis is listed in
Tables \ref{table3.1}, \ref{table3.2} for the adjustment to 1~WU and in Tables
\ref{table3.3}, \ref{table3.4} for 8~Wu, all in percent. They do not add up all to 100~percent,
due to rounding effects and not listing some irreps. The association 
into bands can be guessed,
though for the higher lying states the mixing is important and 
deviates substantially from the internal 
structure of the band heads. This problem is known for a long time \cite{greiner}, making
it difficult to associate a given state to a band.   

In conclusion, several properties and differences are observed:

\begin{itemize}

\item The spin doublets predicted in \cite{bijker} are only partially
reproduced. The doublets at higher energies disappear in the calculation
including the PEP.

\item There is only one $5^-$ state below 25~MeV. The second $4^-$ state
is shifted to larger energies, compared to \cite{bijker}

\item In \cite{bijker} it is claimed that there is no $1^+$ cluster state
possible. However the SACM permits these states and some of them are 
shown in Figure \ref{fig3}. Thus, observed $T=0$ states may very well be 
cluster states!

\item Within the SACM there is an extra $1^-$-band, which is in the
$SU(3)$ limit degenerate to the first one.

\item The association into bands is criticized. While in the oblate top
model it is allowed, as it corresponds also to a molecule with three atoms
where the PEP is of no importance, in a nuclear
cluster system this association is forbidden.

\end{itemize}

Spectroscopic factors for $^8$Be+$\alpha$ were also calculated. For details
of the algebraic spectroscopic factor operator please consult \cite{specfac}.
In \cite{specfac} is is demonstrated that the agreement to microscopic exact
shell model calculations \cite{draayer-specfac} is extremely good. 
Though, the reaction $^8$Be+$\alpha$ $\rightarrow$ $^{12}$C is difficult
or not possible to measure, for completeness the values for some
spectroscopic factors are listed in Table \ref{table4}. Due to a possible
interest of experimental physicists, which look for the $1^+$ states below
25~MeV, also the spectroscopic factors of two $1^+$ states are tabulated and
of the first two states of the second $1^-$-band.

\section{Conclusions}
\label{concl}

We applied the {\it Semimicroscopic Algebraic Cluster Model} \cite{sacm1,sacm2}
to $^{12}$C, as a three-$\alpha$ cluster system. It was shown that it 
suffices to treat $^{12}$C as a $^8$Be+$\alpha$ two cluster system.

Some geometrical considerations were applied and it we showed 
that in its ground state
$^{12}$C must have a triangular structure. 
The main reason for the finite
distance of two $\alpha$ particles is due the 
minimal number of oscillation quanta, required
by the {\it Wildermuth Condition}, i.e. due to the observation of the {\it Pauli 
Exclusion Principle}.

Three calculations were performed, using for comparison a simplified version
of the SACM: The first calculation
was restricted to the $SU(3)$ limit, while in the next two calculations
the experimental $B(E2; 0_2^+ \rightarrow 2_1^+)$ transition was taken
as 1~WU and 8~WU, respectively, where the last value corresponds to the
experimentally observed one. It was shown that this non-zero transition value
indicates a strong mixing of bands, due to the large deformation of the
$^{12}$C nucleus, making it difficult to group the
states into bands. The overall agreement to the experiment is very good 
up to moderately good. 

It was shown that the PEP has the following effects:

\begin{itemize}

\item Some parity doublets reported in \cite{bijker} dissolve, the states
are pushed apart. 

\item There are additional $1^+$ cluster states, forbidden in the oblate
top model. 

\item There is an extra $1^-$-band in the spectrum.

\item The experimental $B(E2)$ and $B(E3)$ transitions can be
described within the SACM.

\item The low lying $5^-$ state, predicted in \cite{bijker}
is not present when the PEP is observed.

\item The association into bands, as done in \cite{bijker},
is criticized.

\end{itemize}

We also mentioned some critics on the use of the cut-off value $N$ as
a parameter. 

There are ways to improve the agreement. One is to use the
multi-channel symmetry \cite{arxiv}, which reduces the number
of free parameters and other possibility is to apply no-core
shell  model calculations as in <\cite{draayer-hoyle}. The $^{12}$C
is not an easy nucleus to describe and sometimes one has to do more than
using a simple model. 

To resume, The {\it Pauli Exclusion Principle} is very important and
cannot simply be ignored, even if the results indicate that the model "works".
The PEP is one of the fundamental principles of nature and cannot be set aside!

\section*{Acknowledgments}
We acknowledge financial support form DGAPA-PAPIIT (IN100315) and
CONACyT (project number 251817). We thank J. Cseh and G. L\'evai
(Hungarian Academy of Sciences) for very useful discussions and
comments.

\end{document}